\documentclass[12pt]{article}
\usepackage{amsfonts}
\usepackage{amssymb}
\usepackage{latexsym}
\usepackage{graphicx}
\usepackage[english]{babel}
\begin{document}
\input epsf

\renewcommand{\theequation}{\arabic{section}.\arabic{equation}}

\def\be{\begin{equation}}
\def\ee{\end{equation}}

\begin{flushright}
\end{flushright}
\vspace{20mm}
\begin{center}
{\LARGE  What is the state of the Early Universe?}\\
\vspace{18mm}
{\bf  Samir D. Mathur\footnote{mathur@mps.ohio-state.edu}}\\
\vspace{8mm}
Department of Physics,\\ The Ohio State University,\\ Columbus,
OH 43210, USA\\
\vspace{4mm}
\end{center}
\vspace{10mm}
\thispagestyle{empty}
\begin{abstract}





As we follow our Universe to early times, we find that matter was crushed to high densities, somewhat similar to the behavior at a black hole singularity. String theory has made progress in explaining the internal structure of black holes, so we would like to extrapolate the ideas learnt from black holes to the early Universe. If we assume that we want the most probable state of the Universe at early times, then we should look at the kind of state that describes a generic black hole. This suggests a definite equation of state for the matter in the early Universe. Quantum effects can stretch across macroscopic distances in black holes, and these might be important in understanding the early Universe as well.
\end{abstract}

\bigskip

\bigskip

{\it [Proceedings of The International Conference on Gravitation and Cosmology - 07, Pune, Dec 2007]}


\newpage
\setcounter{page}{1}
\renewcommand{\theequation}{\arabic{section}.\arabic{equation}}
\section{Introduction}\label{introduction}
\setcounter{equation}{0}

Today the  Universe is filled quite sparsely with matter. But as we follow the Universe to earlier times, we find a higher energy density. At sufficiently early times we must have been in a domain where the energy density was set by the Planck scale and quantum gravity was important. 
We do not understand much about this stage of the Universe, but it is likely that to understand the Universe today we would have to understand it in this Planck era.

There are numerous models that try to fit the observed Cosmological data, starting with some assumptions about the state of the Universe at some `post-Planck' era. Thus we might assume the existence of an `inflaton' field in some special state, and see what evolution we get for the metric and matter. 
While such models are clearly important, they do not address some of the most basic puzzles about the Universe. What determines the initial state of the Universe? Should this state be uniquely determined by some principle, or should there be many possible initial states like in any theory with a Hamiltonian evolution? What imprint, if any,  does quantum gravity leave on the state of the Universe as it exits the Planck era? Should we start with a `hot thermal soup' as the initial state as we do in big-bang models? 

\begin{figure}[ht]
\includegraphics[width=14pc]{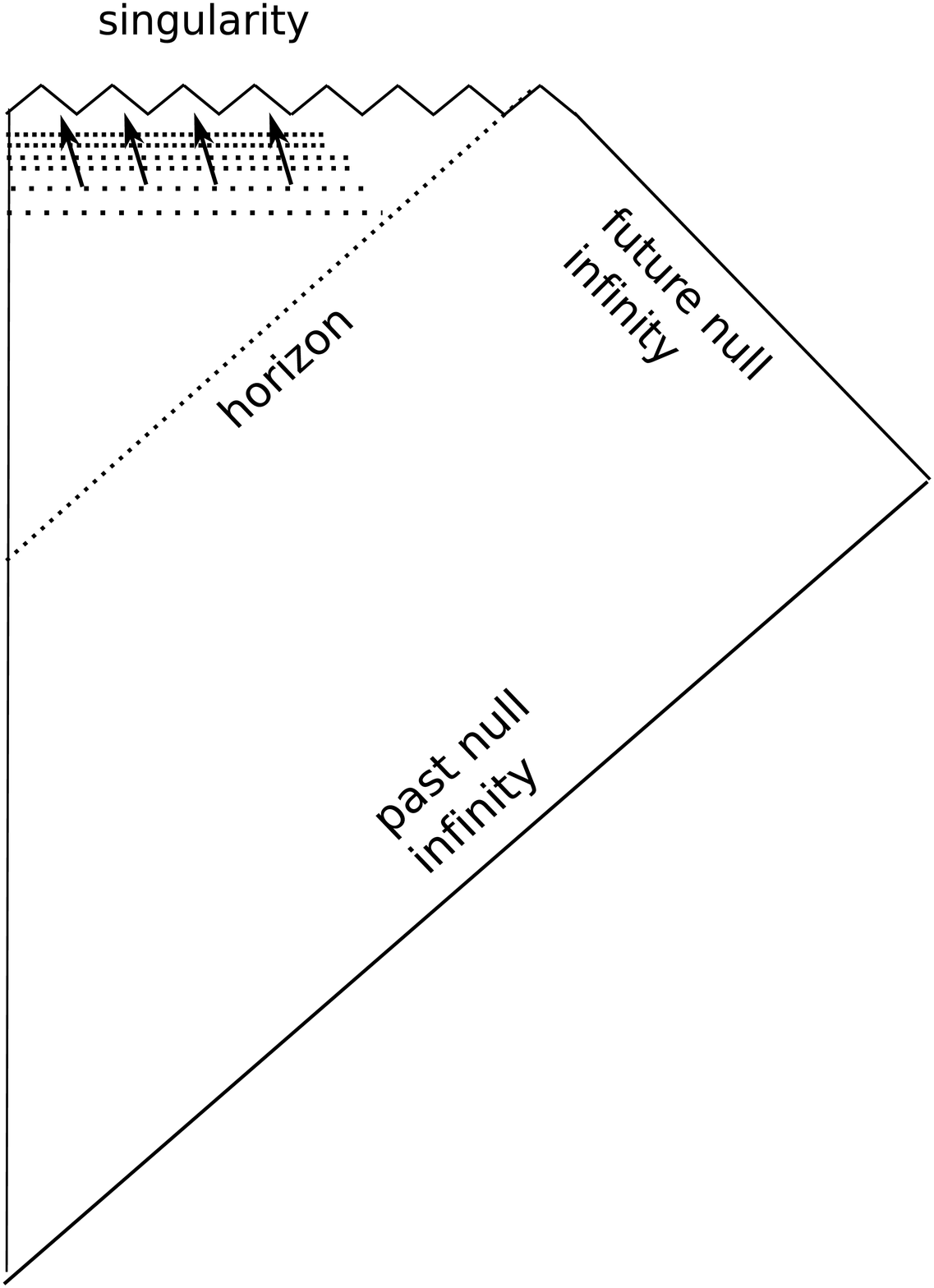}\hspace{2pc}%
\begin{minipage}[b]{14pc}\caption{\label{fone}The Penrose diagram for the classical black hole. Infalling matter approaching the spacelike singularity is represented by  dotted lines. The volume of along spatial slices shrinks as we approach the singularity, sending the density of matter to infinity.}
\end{minipage}
\end{figure}

In the past we did not know how to study the Planck era because we did not have a good theory of quantum gravity. But string theory has now emerged as a consistent quantization of gravity. It has proved remarkably successful in understanding black holes, where we also have matter crushed to high densities, just as in the early Universe. Consider the Penrose diagram for a collapsing black hole drawn in fig.\ref{fone}. As we approach the  spacelike singularity along the slices shown in the figure, the spatial metric becomes small. Thus the slices shrink and the matter density on them  increases, in a manner similar to the increase in density seen as we approach the initial singularity of the Universe.

We should therefore look at what lessons we have learnt from black holes in string theory, and see what they suggest for the early Universe. The hope is that we can reach a set of physical principles that determine the behavior of the Universe, instead of just fitting models to the post-Planck era. 

\section{High density state of the early Universe}

\begin{figure}
\begin{center}
\includegraphics[width=34pc]{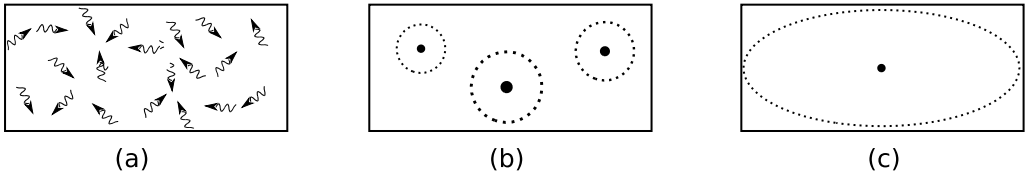}
\end{center}
\caption{\label{ftwo}(a) The early Universe filled with radiation (b) We can get more entropy by collecting the matter into black holes. (c) As we increase the energy, the radius of the black hole will equal the size of the toroidal spatial slice. If we increase the energy even further, what will be the maximal entropy state?  }
\end{figure}

For convenience, we will assume that the spatial slices of our Universe are in the shape of a torus; thus the Universe has a finite volume $V(t)$. As we follow the Universe to earlier times, $V$ becomes small and the energy density becomes large. In the conventional big-bang scenario, the Universe is filled with thermal radiation with temperature $T(t)$; this is depicted in fig.\ref{ftwo}(a). Why do we take such a state? Given a box of volume $V$ and a total energy $E$, the thermal state has the largest value for the entropy $S$, at least in ordinary field theory. Thus we are implicitly assuming that at any time $t$ we want the state of matter to be the one with the highest entropy $S$.

If we allow ourselves to play with gravity, then we might be able to get more entropy by collapsing the matter into black holes. 
Let us keep $V$ fixed and imagine increasing the energy $E$. At very low $E$, the thermal gas has the highest entropy. At larger $E$, we get a higher entropy if the quanta bind together into black holes: we lose the phase space arising from the different allowed positions of these quanta, but gain the  Bekenstein entropy \cite{bek}
\be
S_{bek}={A\over 4G}
\label{qthir}
\ee
that should be associated to each hole. Fig.\ref{ftwo}(b) shows the energy of the Universe distributed among a collection of black holes. We can get an even larger entropy by making just one big black hole. Imagine increasing $E$ till this black hole just barely `fits' into our toroidal Universe (fig.\ref{ftwo}(c)). 
What happens if we increase $E$ still further? In general relativity we are allowed to put  an arbitrarily large energy $E$ in our box of volume $V$; the field equations only imply that there will be  a correspondingly large  $|{dV\over dt}|$. Of course with a large $|{dV\over dt}|$ there may not be a well defined notion of entropy since defining entropy assumes some kind of equilibrium. We are going to ignore this issue and just use equilibrium thermodynamics at each $t$ the same way we do in the standard big bang.

So we have a fairly sharp question that we can ask of string theory: if we take a volume $V$ and keep increasing the energy $E$ in it, then what is the entropy $S(E, V)$ as $E\rightarrow\infty$? Further, what is the nature of the generic state contributing to this entropy?

\section{Black holes in string theory}

To answer this question, let us first look at what string theory has taught us about black holes. 

Black holes present us with two closely related puzzles. Thermodynamic arguments suggest that we should associate the entropy $S_{bek}$ (eq. (\ref{qthir})) with the hole. But then statistical mechanics implies that there should be $e^{S_{bek}}$ states of the hole. The classical black hole solution looks unique (`black holes have no hair'), so where are these different states? Secondly, black holes emit radiation at the rate expected of a warm body. But this radiation is created by the production of particle-antiparticle pairs at the horizon, which is far from the singularity $r=0$ where the matter making the hole went. Thus the radiation carries no information about the initial matter, and when the hole finally evaporates away, we have `lost information' and violated quantum mechanics \cite{hawking}.

\subsection{Entropy of black holes in string theory}

String theory is a rigorous mathematical construction, which has no free parameters. We cannot add or subtract any fields from the theory without making it inconsistent. Thus to make a black hole we must use only the particles present in the theory. 

String theory lives in 9+1 spacetime dimensions. But we see only 3+1 dimensions around us. Thus the extra dimensions must be small compact directions. Let us compactify our spacetime as
\be
M_{9,1}\rightarrow M_{4,1}\times S^1\times T^4
\label{qten}
\ee
We can take a string and wrap it around the $S^1$. From the viewpoint of an observer who cannot resolve the compact directions, this will look like a mass point in the non-compact spacetime, carrying some mass $M$ and also a `winding charge' $Q$ carried by the string. Of course if we want to make a black hole we would like a large mass, so we should take a large number $n_1$ of such strings. It is important that we make a {\it bound} state of these strings, since we want to make one massive hole, and not a collection of different tiny holes. But it is easy to see what such a bound state will look like: we just wind the string $n_1$ times around $S^1$ before joining it back to forming a closed loop.

How much entropy does this object have? The string is an elastic band, and so settles down to have the shortest length for its given winding number. Thus we get a unique state, and so the microscopic entropy is $S_{micro}=\ln 1=0$. 

What about the Bekenstein entropy? The string creates a metric that reflects both its mass and winding charge, and we find that this metric has a horizon which coincides with the singularity: i.e., the area $A_H$ of the horizon is zero and thus $S_{bek}=0$. Thus we get an agreement $S_{micro}=S_{bek}$, but in a somewhat trivial way.

What other objects do we have in our theory? We certainly have the massless graviton, we can let this graviton race around the compact $S^1$. From the viewpoint of someone who cannot resolve the compact directions this again looks like a mass point $M$, but this time carrying `momentum charge' coupling to the Kaluza-Klein gauge field arising from the dimensional reduction along $S^1$. 

To make a massive object we need a large energy, but since we want {\it one} black hole rather than many tiny ones, we should bind all the gravitons into one; i.e. we should just take a single high energy graviton racing along $S^1$. There is only one such state, so the microscopic entropy is $S_{micro}=\ln 1=0$. The metric created by this graviton turns out to have $A_H=0$, and so we again get
$S_{bek}=0=S_{micro}$.

To get something more interesting let us take {\it both}  winding and momentum charges. We need a {\it bound} state of the `multiwound' string and the momentum carrying gravitons. But it is easy to picture what this bound state will look like: the momentum is carried on the string in the form of {\it travelling waves}. But now we  see that there are many ways to put the same total momentum on the  string: we can put all the energy in the lowest harmonic, or some in the first and some in the second and so on. Let us see how to compute the number of partitions among these harmonics.

Let the length of the $S^1$ be $L$. The total momentum charge must have the form
\be
P={2\pi n_p\over L}
\ee
with $n_p$ an integer. 
Let us open up the `multiwound' string to its total length $L_T=n_1 L$. We can write
\be
P={2\pi n_1n_p\over n_1 L}={2\pi n_1n_p\over L_T}
\ee
The lowest harmonic on the string carries energy and momentum $e=p={2\pi\over L_T}$. The $k$th harmonic carries
\be
e=p={2\pi k\over L_T}
\ee
Let there be $m_k$ excitations in the harmonic $k$. We need
\be
\sum_k  k\,  m_k=n_1n_p
\ee
Thus we need to compute the `partitions' of the integer $n_1n_p$. The number of partitions of an integer $N$ grows like the exponential of $\sqrt{N}$. In our case we must also take into account that there are $8$ different transverse directions in which the string can vibrate (the total spacetime dimension is 9+1),
so there are $8$ kinds of bosonic excitations with the same $k$. Further there are $8$ fermionic superpartners of these $8$ bosonic vibrations for the superstring. With all this we find for the number of states
\be
{\cal N} \sim e^{2\sqrt{2}\sqrt{n_1n_p}}
\label{qfourt}
\ee
Thus the microscopic entropy of the bound state carrying winding and momentum charges $n_1n_p$ is \cite{sen}
\be
S_{micro}=\ln {\cal N} =2\sqrt{2}\sqrt{n_1n_p}
\label{qfift}
\ee

What about the Bekenstein entropy of a black hole with these charges? This entropy has been computed in a  closely related  case also carrying  two charges. The charges correspond to D4 and D0 branes, and the spacetime is compactified down to $M_{3,1}$.  To make the gravity solution we have to solve the field equations of the metric and gauge fields. There are two things to note in carrying out this computation. The first is that in string theory the Einstein action $R$ gets higher order corrections of type $R^2$, and it will be these corrections that will make the horizon area nonzero. (These corrections also modify the  Bekenstein entropy  to the `Bekenstein-Wald entropy' \cite{wald}.) The second point is that we solve the resulting field equations with an ansatz that imposes spherical symmetry in the noncompact directions. Such black hole solutions were made  in \cite{dabholkar} and the Bekenstein-Wald entropy computed. The microscopic entropy can  be counted by understanding the structure of the bound state \cite{vafa}, and gives an answer similar to  (\ref{qfift}). We find
\be
S_{micro}=4\pi\sqrt{n_0n_4}=S_{bek}
\ee
We can make more complicated black holes.  For example we take the compactification (\ref{qten}), take the above mentioned winding and momentum charges around the $S^1$ , and add  $n_5$ 5-branes wrapped around $S^1\times T^4$. The geometry will now carry winding, momentum and 5-brane charge, and we can again look for a solution where the metric functions depend on $r$ only. In this case the $R^2$ corrections are subleading, and the Bekenstein entropy can be computed from the Einstein action as ${A_H\over 4G}$. One finds \cite{sv}
\be
S_{micro}=2\pi\sqrt{n_1n_pn_5}=S_{bek}
\ee
All these cases described {\it extremal} black holes, i.e. those with mass=charge. Extremal holes have zero Hawking temperature and do not radiate. We can make non-extremal holes by using the same objects as above but also using their anti-particles. Thus suppose we start with large amounts of  string winding and 5-brane charges $n_1, n_5$, and add a small amount of  extra energy on top of the mass carried by these objects. In the microscopic description the extra energy creates momentum modes $P$ running along the $S^1$ in one direction and modes carrying an equal amount of momentum $\bar P$ running in the {\it opposite} direction along the $S^1$. The entropy can again be compared to the Bekenstein value for the same total mass and charges
and it is found that \cite{callanmalda}
\be
S_{micro}=2\pi\sqrt{n_1n_5}(\sqrt{n_p}+\sqrt{\bar n_p})=S_{bek}
\ee
Now there is more mass than charge, so the state can radiate energy. The $P$ and $\bar P$ excitations collide and annihilate, converting their energy to supergravity quanta that leave the bound state and flow off to infinity. The rate for this process can be computed, and agrees in all it details with the statistical properties of Hawking radiation \cite{radiation}
\be
\Gamma_{micro}=\Gamma_{Hawking}
\ee
We can go further by keeping just one large charge, say the 5-brane charge $n_5$ and adding a little extra energy to make the state nonextremal. It is found that the extra energy on the 5-branes takes the form of  a string loop living inside the 5-branes \cite{maldacenafive}. Low energy radiation from such a bound state is again found to have the same rate as the Hawking radiation from the corresponding black hole solution \cite{km}. 

The extremal black holes are supersymmetric, and in all these cases where we had one or more large charges and a small amount of extra energy the state was `close to supersymmetric'. One can therefore argue that supersymmetry is  protecting the states of the system from getting large energy corrections due to interactions. This might be the reason that   a simple `weak coupling'  computation of microscopic degeneracies is agreeing well with the Bekenstein entropy which needs the gravitational interaction to be strong enough to make a black hole. But, quite remarkably, we can extrapolate the above entropy expressions to a domain where there are {\it no} large charges and we are no longer close to a supersymmetric state. Thus let the state have charges $\hat n_1, \hat n_p, \hat n_5$, and a total energy $E$. The Bekenstein entropy can be written exactly as \cite{hms}
\be
S_{bek}=2\pi(\sqrt{n_1}+\sqrt{\bar n_1})(\sqrt{n_p}+\sqrt{\bar n_p})(\sqrt{n_5}+\sqrt{\bar n_5})
\label{qtw}
\ee
How do we determine the  6 parameters $n_a, \bar n_a$ in this expression? First we have the constraints that the net charge of each type is given by the difference between the charges and anti-charges
\be
\hat n_1=n_1-\bar n_1, ~~~\hat n_p=n_p-\bar n_p, ~~~\hat n_5=n_5-\bar n_5
\ee
Next, for each type of charge there is a mass $m_a$ that corresponds to one unit of that charge sitting in isolation by itself, in flat spacetime. Thus for the string this mass $m_1$ is just the tension $T_1$ (i.e. mass per unit length) of the string times the length $L$ of the $S^1$. Similarly, let the volume of the $T^4$ be $V$ and the tension of the 5-brane be $T_5$. Then we have
\be
m_1=T_1L, ~~~m_p={2\pi\over L}, ~~~m_5=T_5 VL
\ee
The antiparticles have the same energy as the corresponding particles.
We assume that there is no interaction energy between  these objects $n_a, \bar n_a$, though we can prove this only for the supersymmetric case ($\bar n_a=0$) and can argue plausibly for it if we are quite close to the supersymmetric case. We  write
\be
E=(n_1+\bar n_1)m_1+(n_p+\bar n_p)m_p+(n_5+\bar n_5)m_5
\label{qninet}
\ee
Thus we have four constraints on the six variables $n_a, \bar n_a$. We now maximize the expression $S=2\pi(\sqrt{n_1}+\sqrt{\bar n_1})(\sqrt{n_p}+\sqrt{\bar n_p})(\sqrt{n_5}+\sqrt{\bar n_5})$ over the $n_a, \bar n_a$ subject to these four constraints. Computing the value of $S$ at this maximum we find exactly the Bekenstein entropy for a black hole with charges $\hat n_1, \hat n_p, \hat n_5$ and energy $E$. The expression (\ref{qtw}) includes all the special extremal and near-extemal cases that we discussed above, as well as the case of the Schwarzschild hole $\hat n_a=0$.

The charges $n_1, n_p, n_5$ appeared for the compactification (\ref{qten}). We can also compactify down to $3+1$ spacetime dimensions as
\be
M_{9,1}\rightarrow M_{3,1}\times S^1\times \tilde S^1\times T^4
\label{qtenp}
\ee
Now there are more compact cycles, and we will have {\it four} kinds  charged objects that can be wrapped on these cycles. We have string winding charges, momentum modes and 5-branes as before, and in addition Kaluza-Klein monopoles that have $\tilde S^1$ as their nontrivially fibred circle (their metric has no dependence on $S^1\times T^4$). The entropy for extremal, near-extremal and general holes all follow the same pattern as before, with the general case being described by \cite{hlm}
\be
S_{bek}=2\pi(\sqrt{n_1}+\sqrt{\bar n_1})(\sqrt{n_p}+\sqrt{\bar n_p})(\sqrt{n_5}+\sqrt{\bar n_5})(\sqrt{n_{kk}}+\sqrt{\bar n_{kk}})
\label{qtwp}
\ee
The parameters $n_a, \bar n_a$ for the 4-charge case are determined in exactly the way they were determined for the 3-charge case above. 

\section{Fractionation and the fuzzball structure of the black hole interior}

\begin{figure}
\begin{center}
\includegraphics[width=25pc]{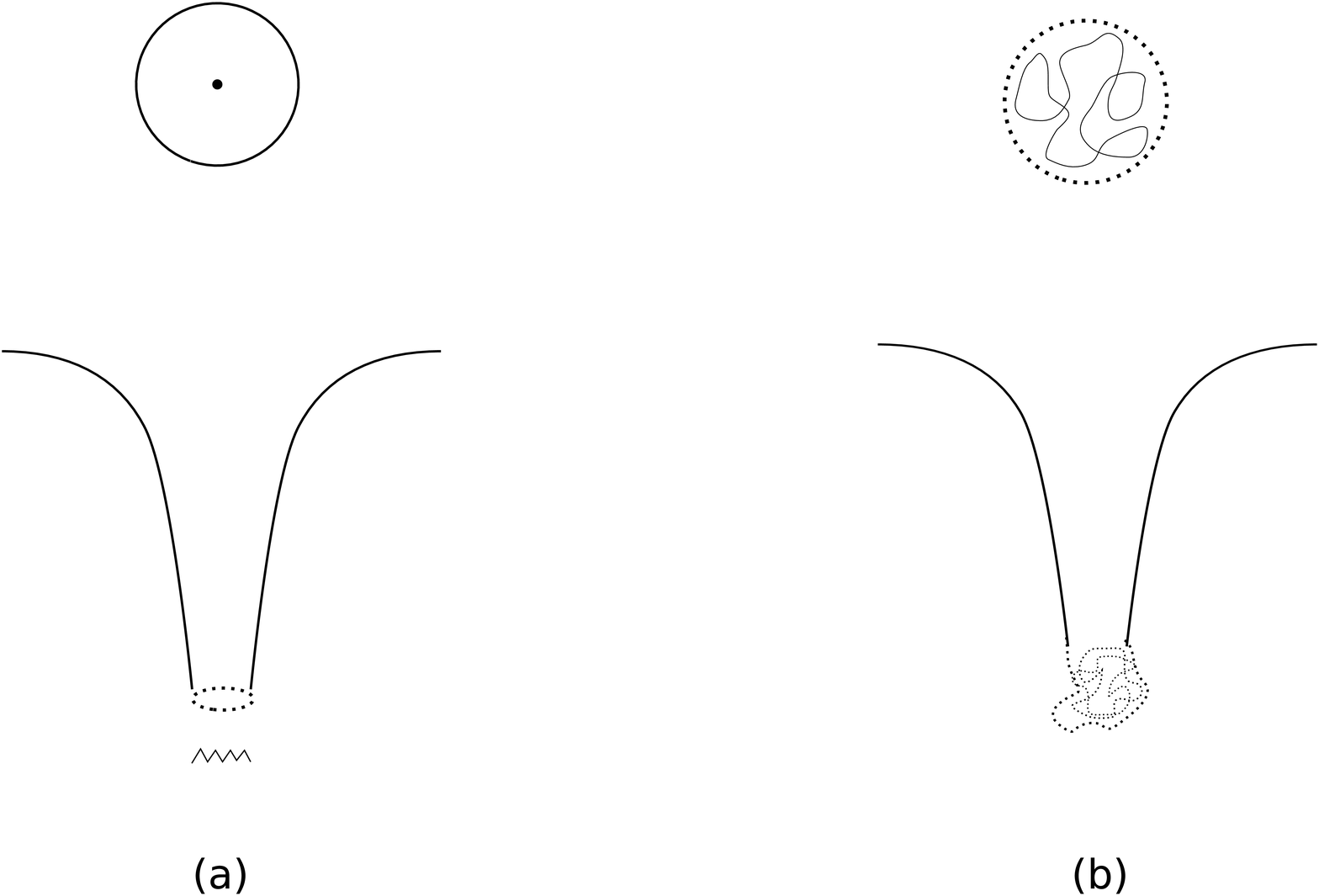}
\end{center}
\caption{\label{fthree}(a) The traditional black hole; if the winding and momentum charges of the string could sit at a central point, then the geometry (pictured below the sketch of the hole) would have a horizon and a singularity. (b) In the actual string theory construction of the bound state the momentum charge causes the string to spread out over a horizon sized transverse region, generating a horizon sized `fuzzball'.}
\end{figure}

All this suggests that we understand how to count states of the hole, but how do we bypass the information paradox? Let us  return to the simple 2-charge extremal  hole made of string winding and momentum charges. The entropy (\ref{qfourt}) comes from the different way in which the string carries the momentum  as travelling waves. It is very important to note that the fundamental string of string theory does not carry any {\it longitudinal} waves; thus all vibrations must be {\it transverse}. Thus in carrying the momentum charge the string departs from its central location $r=0$ and spreads over a certain transverse region, making a `fuzzball'. We depict this in fig.\ref{fthree}. Different  vibration states of the string make different `fuzzballs' \cite{lm4}. No fuzzball has a singularity or a horizon, but if we draw a sphere bounding the typical fuzzball then the area  $A$ of this sphere satisfies
\cite{lm5}
\be
{A\over G}\sim \sqrt{n_1n_p}\sim S_{micro}
\ee
Thus the information of the black hole is distributed throughout the interior of a horizon sized ball. The vicinity of the horizon is not `empty space' as in the traditional classical picture of the hole. Radiation from an excited fuzzball leaves from the surface of the fuzzball, carrying out the information of the state just like radiation from a piece of coal carries information about the structure of the coal.  

Similar constructions have been carried out for the microstates of the more complicated holes discussed above \cite{many}, and in each case the microstate turns out to have no horizon or singularity.  How can this happen, if the natural length scale for quantum gravity is $l_p$, the Planck length? The point is that a black hole is made of a large number of quanta $N$, and so we have to ask if quantum gravity effects range over distances $\sim l_p$ or $\sim N^\alpha l_p$ for some $\alpha$. It turns out that the latter is the case, with $\alpha$ being such that the scale of the nonlocal effects is of order the horizon size, and  the entire interior of the hole turns into a very quantum `fuzzball'.

What effect in string theory is responsible for generating this large length scale? String theory manifests an effect called {\it fractionation} \cite{dmpre}.  Consider a compact circle $S^1$ of length $L$, and let us ask what is the minimum energy that we need to make an excitation, with no net change of the charges. We can take a pair of gravitons in the lowest allowed harmonic, one moving clockwise and one anticlockwise around the $S^1$. This gives an energy gap $(\Delta E)_0={2\pi\over L}+{2\pi\over L}$. But now suppose that we had a string wrapped $n_1$ times on this $S^1$. Then  we can make an excitation by letting the gravitons be travelling waves on the string, and now the lowest allowed energy of vibration is
\be
\Delta E={2\pi\over L_T}+{2\pi\over L_T}={4\pi\over n_1 L}={1\over n_1} (\Delta E)_0
\ee
Thus if the momentum modes $P, \bar P$ are allowed to {\it bind} to $n_1$ units of string winding charge, then they come in {\it fractional} units that are ${1\over n_1}$th  of the normal unit of quantization. This is a simple effect by itself, but in string theory we have dualities that can map any excitation to any other. Thus we can map the $n_1$ strings to $n_1$ 5-branes, and under the duality the fractional $P-\bar P$ pairs become fractional {\it string} pairs \cite{maldasuss}. But a fractional string is a string with tension ${1\over n_1}$ times the string tension $T_1$. Thus for large $n_1$ we generate a very low tension, `floppy' object, which can stretch to large distances. In \cite{emission} it was found that for the extremal 3-charge bound states discussed above the distances to which these floppy objects extend is 
\begin{equation}
D\sim \left [ {(n_1n_pn_5)^{1\over 2} g^2 \alpha'^4\over VL}\right ] ^{1\over 3}\sim R_s
\label{matqqq}
\end{equation}
where $R_s$ is the Schwarzschild radius of the extremal black hole with these charges! Thus string theory naturally generates a length scale that grows with the number of charges in the bound state, and  quantum gravity effects stretch across the entire radius of the hole. Information can therefore leave with the radiation from the surface of the fuzzball, and we resolve the information paradox.

One may still ask the question: what happens to a shell of matter as it collapses through its horizon? This process seems to be well described by classical gravity, and seems to lead to a traditional black hole rather than a `fuzzball'. The point to note is that this collapsing shell gives a time-dependent configuration with very low entropy. We need to wait for the system to `ergodize' and reach a generic state, which we expect to be a quantum, horizon-sized fuzzball. Hawking radiation is a very slow process, so all that we need to have to prevent information loss is that this relaxation to the generic state happen in a time shorter than the Hawking evaporation time. 

How can the `relaxation' occur? The dynamics of fuzzballs is still in its infancy. But we can imagine a tunneling process that leads from  the collapsing shell state to a fuzzball state. The action for this state will be large since we are relating two different macroscopic structures. We can estimate this action to be order  
\be
S_{tunneling}\sim {1\over G}\int R\sqrt{-g}\sim {M^2\over G}
\ee
The resulting tunneling amplitude $\sim e^{-S_{tunneling}}$ is very small, and this smallness is the reason that other macroscopic objects do not change from one state to a completely different state. But for the black hole something special happens: there are $e^{S_{bek}}$ different fuzzball states that we can tunnel to, and this is a very large number. In fact $S_{bek}\sim {M^2\over G}$, so the overall probability for tunneling out of the shell state involves the product
\be
e^{S_{tunneling}}e^{S_{bek}}\sim 1
\ee
Thus with the large entropy of black holes it is possible to encounter phenomena that would not be expected from ordinary macroscopic objects. It would be useful to explore such possibilities in more detail.

\section{Extracting lessons for the early Universe}

Let us now return to our toroidal Universe, and address the question we had raised: if we fix the volume of this toroidal box and take $E\rightarrow\infty$ then what is the most entropic state in the box? A thermal gas has
$S\sim E^{D-1\over D}$, where $D$ is the dimension of spacetime. For large $E$ we can do better with a `string gas', which has $S\sim E$, and it was argued in \cite{bran} that the early Universe was in fact described by such a string gas. We can think of a string gas as a  `2-charge' state. Thus consider an excited string, carrying no net winding or momentum.  One side of the string loop is a `winding mode', while the other side is an `anti-winding mode'. The momentum runs along the string in both directions, making $P-\bar P$ pairs with no net momentum charge.  It can be shown that the energy of the string comes equally from the tension of the string and from the vibrations on the string; thus each of these kinds of energy has value $E/2$. We can then write the entropy of 2-charge states (see (\ref{qfift}),(\ref{qtw})) as $S\sim \sqrt{({E\over 2})({E\over 2})}\sim E$, which is the equation of state of the string gas.

But if we accept the principle that we should look at the state with maximum entropy $S$ for a given $E,V$, then we see that we can do even better if we use {\it three} kinds of charges. With three charges, if we put energy ${E\over 3}$ in each charge, we get an entropy $S\sim E^{3\over 2}$ (see eq. (\ref{qtw})). With four kinds of charges, we will get $S\sim E^2$ (eq. (\ref{qtwp})). 

Can we extend this pattern further? The more compact directions we have, the more the kinds of objects that we can wind around the compact cycles. In the case of the black hole we do not consider more than 6 compact directions because with less than 3+1 noncompact directions  the spacetime is not asymptotically flat. For our Universe {\it all} spatial directions are compact. We will let the spacetime dimension be $D$, though finally we intend to take D=11 to get M theory description of string theory. The spatial directions are compactified as $x_i\sim x_i+L_i$. We can wrap the extended objects of the theory around the cycles of the torus. A $p$-brane wraps $p$ of these cycles, and has a mass $M=T_p \prod_k L_k$ where $T_p$ is the tension of the $p$-brane and the product runs over the wrapped cycles.   Looking at the expressions (\ref{qtw}),(\ref{qtwp}) it is natural to postulate that the entropy will have a form \cite{cm}
\be
S=A_N\prod_{a=1}^N (\sqrt{n_a}+\sqrt{\bar n_a})
\label{eone}
\ee
where $A_N$ is a constant and $n_a, \bar n_a$ are the numbers of branes and antibranes of type $a$.  The Universe must be overall charge neutral since it is compact, so  $n_a=\bar n_a$. Following (\ref{qninet}) we postulate that the total energy $E$ is given simply by the sum over the masses of the various particles and antiparticles
\be
E=\sum_{a=1}^N m_a (n_a+\bar n_a)=2 \sum_{a=1}^N m_a n_a
\label{etwo}
\ee
What can we take for $N$? From the story for black holes we know that $N$ can be at least $4$. The string theory objects appearing in an expression like (\ref{qtw}) are such that any one of these objects forms a supersymmetric bound state with any other object in the expression. Thus we now look at sets of objects in M theory that are `pairwise supersymmetric' but where we now have 10 cycles available to wrap around. It turns out that we can find sets that have upto 9 such different objects, suggesting that we can take $N=9$, but for the computation below we will keep $N$ arbitrary. 

We must now find the $n_a$ by maximizing (\ref{eone}) subject to (\ref{etwo}). This turns out to give
\be
n_a=\bar n_a={E\over 2 N m_a}
\label{ethree}
\ee
and we find that the energy is equipartitioned over the $N$ different species: $E_a=2n_am_a={E\over N}$. 

Next we evolve the Universe with this matter. We take a metric ansatz
\be
ds^2=-dt^2+\sum_i a_i^2(t) dx_idx_i
\label{efour}
\ee
A brane wrapping directions $k=1, \dots p$ has a diagonal stress tensor
\begin{eqnarray}
T^{(p)k}{}_k&=&-T_p ~\prod _{i=p+1}^{D-1}\hat\delta (x_i-\hat x_i), ~~~k=1,\dots p\nonumber \\
T^{(p)k}{}_k&=&0, ~~~~~~~~~~~~~~~~~~~~~~~~~~~~k=p+1, \dots (D-1)
\end{eqnarray}
For any direction $i$ of the Universe let $N_i$ be the number of {\it types} of objects that wrap that cycle. Let $w_i\equiv -{N_i\over N}$. As an example suppose we have M2 branes wrapping the 2-cycles $(1,2),(3,4),(5.6),(7.8),(9,10)$. Then $N=5$, and since each cycle is wrapped by only one kind of brane, $N_i=1$ for each $i$. Thus for this case
\be
w_i=-{1\over 5}, ~~~~~~~~i=1, \dots 10
\label{esix}
\ee

With $n_a$ given by (\ref{ethree}) we find that the total stress-energy tensor is diagonal, with pressures given by \cite{cm}
\be
p_i=-w_i\rho
\label{efive}
\ee
Our task is now to solve Einstein's equations for the metric (\ref{efour}) with matter satisfying (\ref{efive}). Interestingly, this problem can be solved in closed form \cite{cm}. Define the constants
\be
W\equiv \sum_i w_i, ~~~U\equiv \sum_i w_i^2, ~~~K_1\equiv{(D-1-W)\over 2(D-2)},~~~K_2\equiv-{1\over 2}[{1-W\over D-2} W+U], ~~~\delta_k\equiv{1\over 2}[{1-W\over D-2}+w_k]
\ee
Then
\be
a_k=C_k (\tau-r_1)^{2(\delta_k r_1+f_k)\over (K1+K_2)(r_1-r_2)}
(\tau-r_2)^{2(\delta_k r_2+f_k)\over (K1+K_2)(r_1-r_2)}
\ee
where $\tau$ is an auxiliary time variable defined through
\be
t-t_0=A_4\int_0^\tau d\tau(\tau'-r_1)^{{2(-r_1 K_2 + A_2)\over (K_1+K_2)(r_1-r_2)}}(\tau'-r_2)^{{2(-r_2 K_2 + A_2)\over (K_1+K_2)(r_1-r_2)}}
\ee
Note that this expression is an incomplete Beta function $B_x(p,q)=\int_0^x s^{p-1}(1-s)^{q-1} ds$. Here $r_1, r_2$ are defined through
\be
-(K_1+K_2){\tau^2\over 2}+(A_2-A_1)\tau +A_3=-{(K_1+K_2)\over 2}(\tau-r_1)(\tau-r_2)
\ee 
$A_1,A_2, A_3, A_4$ are arbitrary constants, and the constants $f_k$ are subject to $\sum_k f_k=A_1$. 

\begin{figure}[ht]
\includegraphics[width=14pc]{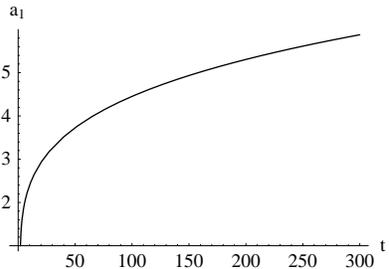}\hspace{2pc}%
\begin{minipage}[b]{14pc}\caption{\label{ffour}The evolution of the scale factor $a_1$ for the case where the brane content is described by (\ref{esix}), and all directions $x_i$ are given the same initial length and rate of increase.}
\end{minipage}
\end{figure}

For the case (\ref{esix}), if we take all $a_i, \dot a_i$ equal at an initial time $t_i$, then we get the evolution given in fig.\ref{ffour}.

\section{The nature of the maximal entropic state}

Black holes are expected to be highly entropic objects, and we have seen that string theory explains this entropy through the behavior of bound states made out of the objects in the theory. We have then extrapolated the form of this entropy to get an expression for the entropy of the Universe (eq. (\ref{eone})), and then solved for the evolution with this equation of state. Let us now discuss the physical nature of the state that has this entropy.

Branes wrapping cycles have been used before in Cosmology. The `string gas' of \cite{bran} was generalized to a `brane gas' in \cite{branp}. It is important to note that our state is {\it not} the same as the state assumed for a brane gas. In a brane gas we imagine a collection of branes that float in the Universe, interacting with each other only when they happen to intersect. The branes can carry vibrations on their surface.  Having a brane costs a certain energy per unit brane area, but if we have enough vibrations on this area then the entropy of these vibrations is such that the total free energy cost of creating the brane becomes zero. This gives a `Hagedorn phase', which has entropy $S\sim E$, whatever be the type of brane that we are considering.

The state that {\it we} are considering is very different. Imagine adding branes to the volume $V$ until they are squeezed so close to each other that they do not float around as independent objects; instead they are forced to form a bound state. Further, we have different species of branes, and when these bind they undergo the phenomenon of `fractionation' discussed above. Recall that when we add a momentum mode to a bound state of $n_1$ strings, we get $n_1$ `fractional' units of momentum. Thus with $n_p$ units of momentum we get $n_1n_p$ fractional objects. These objects can `group' themselves into various subsets, and each such grouping gives a quantum state for the system. The number of such groupings is the number of partitions of $n_1n_p$, which gives an entropy $S\sim \sqrt {n_1n_p}$. With $N$ kinds of charges, and a total energy $E$, we will have $n_a\sim E$ charges of each kind, and a total entropy
$S\sim \sqrt{n_1n_2\dots n_N}\sim E^{N\over 2}$. For large $E$ (and $N>2$) this entropy  is much larger than the Hagedorn entropy $S\sim E$ for the `brane gas'. To summarize, in our state, the energy of the Universe went to making branes and antibranes, which then fused with each other to make `fractional' brane units;  these fractional units then gave an enormous entropy by the same mechanism which gives the entropy to black holes. We will call this state of the Universe the `fractional brane state'.

\begin{figure}
\begin{center}
\includegraphics[width=25pc]{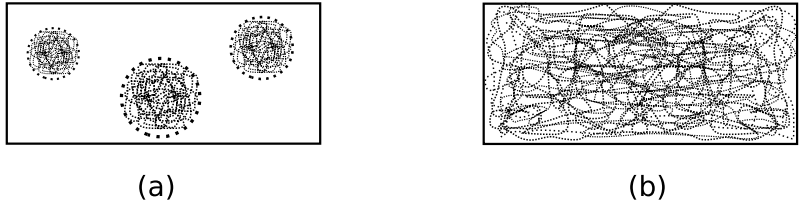}
\end{center}
\caption{\label{ffive}(a) The Universe depicted in fig.\ref{ftwo}(b), with the black hole interiors replaced with their quantum fuzzball states. (b) As we increase the energy $E$, the most entropic state becomes one where the quantum fuzz fills the entire Universe.}
\end{figure}

It is interesting to consider the nature of this state in the context of fig.\ref{ftwo}. In fig.\ref{ftwo}(b) we had sketched the  black holes  in the traditional picture, with no matter outside the central singularity. In fig.\ref{ffive}(a) we sketch the black holes with their interiors described  by  `fuzzball' states.   In fig.\ref{ftwo} it was unclear what would happen if we had more mass than would fit into the hole in fig.\ref{ftwo}(c). But 
there is no such problem now: as we keep increasing the energy we just get  fill the entire Universe with more dense `fuzz' (fig.\ref{ffour}(b)), with the entropy given by the same principle (\ref{eone}) as the one which gave the entropy of the black holes.  

For typical choices of $w_i$ the evolution gives power law type evolution $a_k\sim t^\alpha$ with $\alpha<1$, which is not the kind needed for inflation. Is there some way to account for the observed homogeneity of the Universe at large length scales? Recall that in black holes the quantum effects of these fractional branes stretched across distances of the order of the horizon radius.  In the Universe these nonlocal effects will stretch across the entire torus, thus correlating the physics of any point with any other point. Perhaps such effects may give us an alternative to inflation.

Our assumption at the start was that we should look for the maximal entropy state of the Universe. It may appear that if we start with a maximal entropy state, then the entropy cannot increase further, which would contradict observations today.  In fact the entropy {\it can} increase. To understand this, note that we first fix a volume $V$ and then ask for the state in this box with maximum $S$ for given $E$. But after some time we have a {\it different} value for $V$. The entropy in this larger box can be {\it more} than $S$. As a simple example of this, consider a gas in a cylinder, with a piston at one end. For a given volume $V$ of the cylinder, the gas is in its maximal entropy state. If we adiabatically increase $V$, then $S$ will not change. But if we {\it suddenly} increase $V$, then the entropy {\it will} increase.

A similar phenomenon is expected to happen for the Universe. In the computations discussed above, we have assumed a adiabatic expansion, and $S$ is constant. But in reality the Universe is evolving at a certain rate, and it is plausible that we do not maintain thermodynamic equilibrium of the fractional brane state at all times. Thus if one cycle of the Universe becomes too long, then branes wrapping this cycle become stretched a lot and become very massive. It can then be entropically advantageous to remove such branes from the set of excitations, and put the energy into other branes. If the evolution is too rapid, such changes would not happen adiabatically, and we would get a rise in entropy \cite{cmp}. It is not easy for fractional branes to annihilate, since each fractional brane must find an annihilation  partner `fractionated the same way'. This slow annihilation in the case of black holes had reproduced exactly the rate of Hawking radiation from the hole \cite{radiation}.  It would be interesting to investigate whether the fractional branes in the Universe can annihilate during the Planckian era, or whether they might persist to much later times; if they exist today they might account for dark matter/dark energy.

To close, let us review the philosophy that we are pursuing.  The early Universe presents us with a host of unresolved questions. Should the initial state be arbitrary or fixed by some abstract principle? If it is arbitrary, is there a reason that it should be a generic, `high entropy' state? If we have a `fractional brane state' then how many quanta of the matter  should be in a `bound state' with each other? In the analysis above we had taken a finite Universe and let {\it all} the quanta be bound to each other, but more generally the number of quanta that are bound will determine how far the nonlocal effects of quantum gravity extend. 

But while we do not understand how such questions are to be answered, string theory {\it can} give concrete predictions once we assume some answers to these philosophical questions.\footnote{For some other approaches to Cosmology using string tools,  see \cite{manyrefs}, and for some related work with fractional branes see \cite{kalyan}.} We have seen above that if we take a finite Universe and demand that we be in the state with maximal entropy then string theory suggests a unique equation of state. While we did not {\it prove}  (\ref{eone})  for $N>4$,  this is something that can be  checked within string theory; in any case, it appears to be a well defined question to ask what {\it is} the maximal entropy state for given $E,V$.  Using string theory in this way and comparing to observations we can hope to arrive at an understanding of the fundamental questions mentioned above. This is something that we may not be able to do if we simply try  to fit data to phenomenological models of Cosmology.


\section*{Acknowledgements}

I am grateful to Steve Avery, Borun Chowdhury and Jeremy Michelson for many helpful comments on this manuscript. This work was supported in part by DOE grant DE-FG02-91ER-40690.



\begin{thebibliography}{99}

\bibitem{bek}
  J.~D.~Bekenstein,
  Phys.\ Rev.\  D {\bf 7}, 2333 (1973).

\bibitem{hawking}
  S.~W.~Hawking,
  Commun.\ Math.\ Phys.\  {\bf 43}, 199 (1975)
  [Erratum-ibid.\  {\bf 46}, 206 (1976)];
  S.~W.~Hawking,
  Phys.\ Rev.\  D {\bf 14}, 2460 (1976).
  
\bibitem{sen}
L.~Susskind,
arXiv:hep-th/9309145;
J.~G.~Russo and L.~Susskind,
Nucl.\ Phys.\ B {\bf 437}, 611 (1995)
[arXiv:hep-th/9405117];
%
A.~Sen,
Nucl.\ Phys.\ B {\bf 440}, 421 (1995)
[arXiv:hep-th/9411187];
A.~Sen,
Mod.\ Phys.\ Lett.\ A {\bf 10}, 2081 (1995)
[arXiv:hep-th/9504147];

\bibitem{wald}
  R.~M.~Wald,
  Phys.\ Rev.\  D {\bf 48}, 3427 (1993)
  [arXiv:gr-qc/9307038].


  \bibitem{dabholkar}
  A.~Dabholkar,
  Phys.\ Rev.\ Lett.\  {\bf 94}, 241301 (2005)
  [arXiv:hep-th/0409148].
  
  \bibitem{vafa}
  C.~Vafa,
  Nucl.\ Phys.\  B {\bf 463}, 435 (1996)
  [arXiv:hep-th/9512078].
  
\bibitem{sv}
A.~Strominger and C.~Vafa,
Phys.\ Lett.\ B {\bf 379}, 99 (1996)
[arXiv:hep-th/9601029];
%

\bibitem{callanmalda}
C.~G.~.~Callan and J.~M.~Maldacena,
Nucl.\ Phys.\ B {\bf 472}, 591 (1996)
[arXiv:hep-th/9602043].
%

\bibitem{radiation}
S.~R.~Das and S.~D.~Mathur,
Nucl.\ Phys.\ B {\bf 478}, 561 (1996)
[arXiv:hep-th/9606185];
  J.~M.~Maldacena and A.~Strominger,
  Phys.\ Rev.\  D {\bf 55}, 861 (1997)
  [arXiv:hep-th/9609026].
  
\bibitem{maldacenafive}
  J.~M.~Maldacena,
  Nucl.\ Phys.\  B {\bf 477}, 168 (1996)
  [arXiv:hep-th/9605016].
  
\bibitem{km}
  I.~R.~Klebanov and S.~D.~Mathur,
  Nucl.\ Phys.\  B {\bf 500}, 115 (1997)
  [arXiv:hep-th/9701187].

\bibitem{hms}
  G.~T.~Horowitz, J.~M.~Maldacena and A.~Strominger,
  Phys.\ Lett.\ B {\bf 383}, 151 (1996)
  [arXiv:hep-th/9603109].
  
  
  
  
  \bibitem{hlm}
  G.~T.~Horowitz, D.~A.~Lowe and J.~M.~Maldacena,
  Phys.\ Rev.\ Lett.\  {\bf 77}, 430 (1996)
  [arXiv:hep-th/9603195];
  C.~V.~Johnson, R.~R.~Khuri and R.~C.~Myers,
  Phys.\ Lett.\ B {\bf 378}, 78 (1996)
  [arXiv:hep-th/9603061].

\bibitem{lm4}
O.~Lunin and S.~D.~Mathur,
Nucl.\ Phys.\ B {\bf 623}, 342 (2002)
[arXiv:hep-th/0109154].
%
\bibitem{lm5}
O.~Lunin and S.~D.~Mathur,
Phys.\ Rev.\ Lett.\  {\bf 88}, 211303 (2002)
[arXiv:hep-th/0202072].
%
\bibitem{many}
 S.~Giusto, S.~D.~Mathur and A.~Saxena,
  Nucl.\ Phys.\  B {\bf 710}, 425 (2005)
  [arXiv:hep-th/0406103];
  O.~Lunin,
JHEP {\bf 0404}, 054 (2004)
[arXiv:hep-th/0404006].
I.~Bena and N.~P.~Warner,
  arXiv:hep-th/0701216;
 I.~Bena and N.~P.~Warner,
  Adv.\ Theor.\ Math.\ Phys.\  {\bf 9}, 667 (2005)
  [arXiv:hep-th/0408106];
  V.~Balasubramanian, E.~G.~Gimon and T.~S.~Levi,
  arXiv:hep-th/0606118;
 I.~Kanitscheider, K.~Skenderis and M.~Taylor,
  arXiv:0704.0690 [hep-th];
 V.~Jejjala, O.~Madden, S.~F.~Ross and G.~Titchener,
  Phys.\ Rev.\  D {\bf 71}, 124030 (2005)
  [arXiv:hep-th/0504181];
 V.~Cardoso, O.~J.~C.~Dias, J.~L.~Hovdebo and R.~C.~Myers,
  Phys.\ Rev.\  D {\bf 73}, 064031 (2006)
  [arXiv:hep-th/0512277].
  B.~D.~Chowdhury and S.~D.~Mathur,
  arXiv:0711.4817 [hep-th].


\bibitem{dmpre}
  S.~R.~Das and S.~D.~Mathur,
  Phys.\ Lett.\  B {\bf 375}, 103 (1996)
  [arXiv:hep-th/9601152].
  
   
\bibitem{maldasuss}
J.~M.~Maldacena and L.~Susskind,
Nucl.\ Phys.\ B {\bf 475}, 679 (1996)
[arXiv:hep-th/9604042].

  
\bibitem{emission}
S.~D.~Mathur,
Nucl.\ Phys.\ B {\bf 529}, 295 (1998)
[arXiv:hep-th/9706151].
%

\bibitem{bran}
  R.~H.~Brandenberger and C.~Vafa,
  Nucl.\ Phys.\ B {\bf 316}, 391 (1989);
  R.~H.~Brandenberger, A.~Nayeri, S.~P.~Patil and C.~Vafa,
  arXiv:hep-th/0608121;
  N.~Deo, S.~Jain, O.~Narayan and C.~I.~Tan,
  Phys.\ Rev.\ D {\bf 45}, 3641 (1992);
 M.~Sakellariadou,
  Nucl.\ Phys.\ B {\bf 468}, 319 (1996)
  [arXiv:hep-th/9511075].



\bibitem{branp}
  S.~Alexander, R.~H.~Brandenberger and D.~Easson,
  Phys.\ Rev.\ D {\bf 62}, 103509 (2000)
  [arXiv:hep-th/0005212];
  R.~Easther, B.~R.~Greene, M.~G.~Jackson and D.~Kabat,
  Phys.\ Rev.\ D {\bf 67}, 123501 (2003)
  [arXiv:hep-th/0211124];
  R.~Easther, B.~R.~Greene, M.~G.~Jackson and D.~Kabat,
  JCAP {\bf 0401}, 006 (2004)
  [arXiv:hep-th/0307233];
  R.~Brandenberger, D.~A.~Easson and A.~Mazumdar,
  Phys.\ Rev.\  D {\bf 69}, 083502 (2004)
  [arXiv:hep-th/0307043];
  T.~Battefeld and S.~Watson,
  Rev.\ Mod.\ Phys.\  {\bf 78}, 435 (2006)
  [arXiv:hep-th/0510022];
  S.~Watson,
  Phys.\ Rev.\  D {\bf 70}, 066005 (2004)
  [arXiv:hep-th/0404177].

  \bibitem{cm}
 B.~D.~Chowdhury and S.~D.~Mathur,
  Class.\ Quant.\ Grav.\  {\bf 24}, 2689 (2007)
  [arXiv:hep-th/0611330].
  
  \bibitem{cmp}
  B.~D.~Chowdhury and S.~D.~Mathur, {\it unpublished}.
  
  \bibitem{manyrefs}
    G.~R.~Dvali, G.~Gabadadze and M.~Porrati,
  Phys.\ Lett.\ B {\bf 485}, 208 (2000)
  [arXiv:hep-th/0005016];
  G.~Shiu and B.~Underwood,
  arXiv:hep-th/0610151;
  H.~Firouzjahi and S.~H.~H.~Tye,
  Phys.\ Lett.\ B {\bf 584}, 147 (2004)
  [arXiv:hep-th/0312020];
   S.~Kachru, R.~Kallosh, A.~Linde, J.~M.~Maldacena, L.~McAllister and S.~P.~Trivedi,
  JCAP {\bf 0310}, 013 (2003)
  [arXiv:hep-th/0308055];
  S.~Kachru, R.~Kallosh, A.~Linde and S.~P.~Trivedi,
  Phys.\ Rev.\ D {\bf 68}, 046005 (2003)
  [arXiv:hep-th/0301240];
B.~Craps, S.~Sethi and E.~P.~Verlinde,
  JHEP {\bf 0510}, 005 (2005)
  [arXiv:hep-th/0506180];
 C.~S.~Chu and P.~M.~Ho,
  JHEP {\bf 0604}, 013 (2006)
  [arXiv:hep-th/0602054];
  S.~R.~Das, J.~Michelson, K.~Narayan and S.~P.~Trivedi,
  Phys.\ Rev.\ D {\bf 74}, 026002 (2006)
  [arXiv:hep-th/0602107];
   H.~Ooguri, C.~Vafa and E.~P.~Verlinde,
  Lett.\ Math.\ Phys.\  {\bf 74}, 311 (2005)
  [arXiv:hep-th/0502211];
T.~Hertog and G.~T.~Horowitz,
  JHEP {\bf 0504}, 005 (2005)
  [arXiv:hep-th/0503071];
  A.~Karch and L.~Randall,
  Phys.\ Rev.\ Lett.\  {\bf 95}, 161601 (2005)
  [arXiv:hep-th/0506053];
   S.~Kalyana Rama,
  Phys.\ Lett.\ B {\bf 638}, 100 (2006)
  [arXiv:hep-th/0603216];
   J.~Brown, O.~J.~Ganor and C.~Helfgott,
  JHEP {\bf 0408}, 063 (2004)
  [arXiv:hep-th/0401053];
  E.~Keski-Vakkuri and M.~S.~Sloth,
  JCAP {\bf 0308}, 001 (2003)
  [arXiv:hep-th/0306070];
 R.~Durrer, M.~Kunz and M.~Sakellariadou,
  Phys.\ Lett.\ B {\bf 614}, 125 (2005)
  [arXiv:hep-th/0501163];
T.~Biswas,
  arXiv:0801.1315 [hep-th];
  G.~Minton and V.~Sahakian,
  Phys.\ Rev.\  D {\bf 77}, 026008 (2008)
  [arXiv:0707.3786 [hep-th]];
   K.~Enqvist, N.~Jokela, E.~Keski-Vakkuri and L.~Mether,
  JCAP {\bf 0710}, 001 (2007)
  [arXiv:0706.2294 [hep-th]].
  
  \bibitem{kalyan}
  S.~Kalyana Rama,
  Phys.\ Lett.\  B {\bf 656}, 226 (2007)
  [arXiv:0707.1421 [hep-th]];
  S.~Kalyana Rama,
  Gen.\ Rel.\ Grav.\  {\bf 39}, 1773 (2007)
  [arXiv:hep-th/0702202].
  
\end{thebibliography}
\end{document}